# Extrinsic spin angular momentum carried by plasmonic vortex in terahertz cylindrical corrugated waveguide

Y. Annaka[1*] and K. Ogura[2]

[1]Faculty of Engineering, Niigata University, Niigata, 950-2181, Japan
[2]Graduate school of Science and Technology, Niigata University, Niigata 950-2181, Japan
*annaka@eng.niigata-u.ac.jp

An electromagnetic surface wave carrying orbital angular momentum (AM) is called plasmonic vortex that also carries spin AM originating from evanescent field. In this study, spin and orbital AM of plasmonic vortex in a terahertz cylindrical corrugated waveguide (CCW) are analyzed. We report the extrinsic spin AM of the plasmonic vortex for the first time. The extrinsic spin AM is produced by the rotation of the polarization of the plasmonic vortex whose rotation axis is positioned at the vicinity of the CCW wall. The extrinsic spin AM depends on the intrinsic orbital AM and radius of CCW. The plasmonic vortex with sufficiently large orbital AM carries large spin AM ($> \sigma\hbar$) per photon. This large extrinsic spin AM expands degree of freedom of the spin AM and would be applied to various terahertz technologies.

## I. INTRODUCTION

Light carries angular momentum (AM) which is attractive for various applications including communication technology and optical tweezer [1-8]. AM can be decomposed into orbital and spin parts. The orbital AM of light produced by azimuth phase distribution of a phase factor exp($im\theta$) with topological charge $m$. The orbital AM is classified into intrinsic and extrinsic AM although only intrinsic spin AM is known [2]. The intrinsic orbital AM is quantized as $m\hbar$ per photon whereas the extrinsic orbital AM depends on choice of the origin of the coordinate. The optical spin AM is produced by helicity $\sigma = \pm1$, which represents direction of a circular polarization, and is equivalent to $\sigma\hbar$ per photon. Difference of intrinsic and extrinsic orbital AM appears in optical spin-orbit coupling [9, 10]. The optical spin-orbit coupling finds utility controlling orbital AM of light [11-14]. A spin-orbit coupling takes place when circularly polarized light propagates into an anisotropic inhomogeneous media with geometric phase [11, 15-17]. Y. Gorodetski et al. have reported excitation of electromagnetic (EM) vortex surface wave with spin-dependent topological charge in an anisotropic inhomogeneous plasmonic structure [12, 14]. The spin AM of light plays important roles in controlling the orbital AM. In this context, spin AM of EM surface wave has considerable features which are pointed out in recent theoretical studies [18-20]. The spin AM of the surface wave originates from its evanescent field that exhibits extraordinary features such as transverse spin AM and helicity-independency.

Spin and orbital AM of EM wave have been formulated [18-24]. Sum of the spin and orbital AM is the total AM $\boldsymbol{J}$ that is expressed as $\boldsymbol{J} = \boldsymbol{r} \times \boldsymbol{p}$ with position vector $\boldsymbol{r}$ and momentum density vector $\boldsymbol{p} = \boldsymbol{P}/c^2$. $\boldsymbol{P}$ is Poynting vector and $c$ is the speed of light. The momentum density vector $\boldsymbol{p}$ is decomposed into spin and orbital components as $\boldsymbol{p} = \boldsymbol{p}_o + \boldsymbol{p}_s$ [21]. The orbital AM density is expressed as $\boldsymbol{L} = \boldsymbol{r} \times \boldsymbol{p_o}$, and the spin AM density is $\boldsymbol{S}$ where $\boldsymbol{p_s} = \nabla \times \boldsymbol{S}/2$. The spin AM is generally intrinsic value which excludes the position vector. The formulas of AM are valid for the EM surface wave.

In this study, we theoretically analyze orbital and spin AM of EM surface wave in terahertz cylindrical corrugated waveguide (CCW), and show surprising result: EM surface wave in CCW carries extrinsic spin AM. The CCW is often applied to intense terahertz generation [25-27]. A multimode of terahertz surface wave can be excited up to topological charge $m = 30$ in the 0.1-THz CCW [28]. The surface wave carrying orbital AM is called plasmonic vortex. We analyze the spin AM of the plasmonic vortex varying the topological charge $m$, and show the extraordinary feature of the spin AM of the plasmonic vortex.

## II. FIELD PROPERTY OF PLASMONIC VORTEX

The schematic of the 0.1-THz CCW is shown in Fig. 1. The CCW has a cylindrical periodic structure on the wall. The periodic structure enables to form EM surface wave in terahertz frequency [29, 30]. The radius of the structure periodically changes along the axial direction. The size parameters are periodic length $z_0 = 0.5$ mm, corrugation amplitude $h = 0.3$ mm, and groove width $d = 0.3$ mm. The radiuses $R = 10.0$ mm and 14.85 mm are used in this calculation. In accordance with Floquet theorem, EM wave in the periodic structure possesses Floquet's spatial harmonics, and the EM fields are expressed as the sum of all spatial harmonics. The field vectors $E$ and $B$ are the sum of Floquet's spatial harmonics as follows.

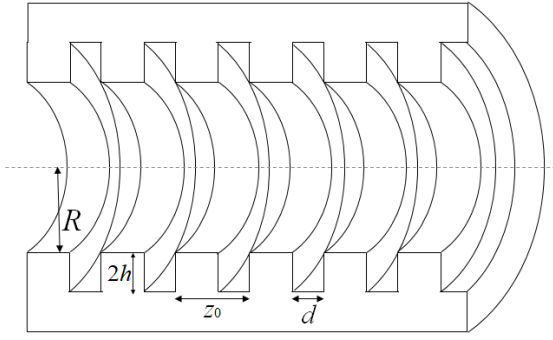

Fig. 1 Cross-sectional view of CCW. CCW has a rectangular periodic structure on wall in order to form terahertz surface wave.

$$E = \sum_{q=-\infty}^{\infty} E_q, B = \sum_{q=-\infty}^{\infty} B_q \quad (1)$$

We use cylindrical coordinate in this study. $E_q = (E_{qr}, E_{q\theta}, E_{qz})$ and $B_q = (B_{qr}, B_{q\theta}, B_{qz})$ are respectively the $q$th Floquet's harmonic of electric and magnetic field vectors with field components of $E_{qr}$, $E_{q\theta}$, $E_{qz}$, $B_{qr}$, $B_{q\theta}$, and $B_{qz}$. The axial component of the $q$th harmonic of the electric and magnetic fields are expressed as bellow.

$$E_{qz} = A_{Eq} I_m(x_q r) \exp(ik_q z + im\theta - i\omega t)$$
$$B_{qz} = \frac{i}{c} A_{Bq} I_m(x_q r) \exp(ik_q z + im\theta - i\omega t) \quad (2)$$

Here, $A_{Eq}$ and $A_{Bq}$ are respectively amplitudes of the $q$th harmonics of electric and magnetic fields, $I_m$ is the $m$th order of the modified Bessel function of first kind, $\omega$ is the angular frequency, $k_z$ is the axial wavenumber, $x_q = (k_q^2 - (\omega/c)^2)^{0.5}$, $k_q = k_z + qk_0$, and $k_0 = 2\pi/z_0$. The other field components are expressed by axial electric and magnetic fields $E_{qz}$ and $B_{qz}$ as follows.

$$E_{qr} = \frac{1}{k_q^2 - \frac{\omega^2}{c^2}} \left( -ik_q \frac{\partial E_{qz}}{\partial r} + \frac{m\omega}{r} B_{qz} \right)$$
$$E_{q\theta} = \frac{1}{k_q^2 - \frac{\omega^2}{c^2}} \left( i\omega \frac{\partial B_{qz}}{\partial r} + \frac{mk_q}{r} E_{qz} \right)$$
$$B_{qr} = \frac{1}{k_q^2 - \frac{\omega^2}{c^2}} \left( -ik_q \frac{\partial B_{qz}}{\partial r} - \frac{m\omega}{c^2 r} E_{qz} \right)$$
$$B_{q\theta} = \frac{1}{k_q^2 - \frac{\omega^2}{c^2}} \left( -\frac{i\omega}{c^2} \frac{\partial E_{qz}}{\partial r} + \frac{mk_q}{r} B_{qz} \right) \quad (3)$$

These fields satisfy equation $\mathbf{k} \cdot \mathbf{E} = \mathbf{k} \cdot \mathbf{B} = 0$ with wave vector $\mathbf{k} = (k_r, m/R, k_z)$. The EM wave in CCW becomes surface wave when radial wavenumber $k_r = ((\omega/c)^2 - k_z^2 - (m/R)^2)^{0.5}$ is of imaginary value. The EM wave in CCW generally forms TM-polarized or TE-polarized waveguide mode. The EM surface wave with $m = 0$ becomes TM-polarized mode. Then the axial magnetic field $B_z$ is equal to zero. On the other hand, the plasmonic vortex with $|m| > 0$ forms hybrid polarized mode of TM and TE. Then all field components are non-zero complex values. Fig. 2 shows real parts of the fields in $r$-$\theta$ plane with $m = 2$ for (a) Re($E_r$), (b) Re($E_\theta$), and (c) Re($E_z$). Here, radius $R = 10.0$ mm, frequency is 103.14 GHz, and axial wavenumber $k_z = \pi/z_0 = 62.83$ cm$^{-1}$. The frequency and wavenumber are determined by the size parameter of the periodic structure, and are calculated by using the field-matching method developed in Ref. [27]. In the method, we consider the fields inside and outside the groove, and solve Maxwell's equations with the boundary condition at the corrugated wall. At this frequency and wavenumber, the phase velocity $\omega/k_z$ is much slower than the speed of light, and then the EM wave in CCW becomes a bounded surface wave [31]. The fields are evanescent being confined to the wall. The field has azimuthal phase distribution owing to phase factor $\exp(im\theta)$. The phase is rotated along the azimuthal direction as the plasmonic vortex propagates. The rotation of the phase produces orbital AM in the direction of the rotation axis. On the other hand, spin AM is produced by a rotation of the polarization. The polarization is rotated owing to the phase differences among the field components. In Fig. 2, the phase difference between $E_r$ and $E_z$ is zero whereas the differences between $E_\theta$ and other components is $\pi/2$. The rotation axis of the spin AM is positioned at each point of the vicinity of the cylindrical corrugated wall while the axis of the orbital AM is at the center of the CCW as shown in Fig. 3. The rotation of the polarization whose axis is off-center of the plasmonic vortex would produce extrinsic spin AM.

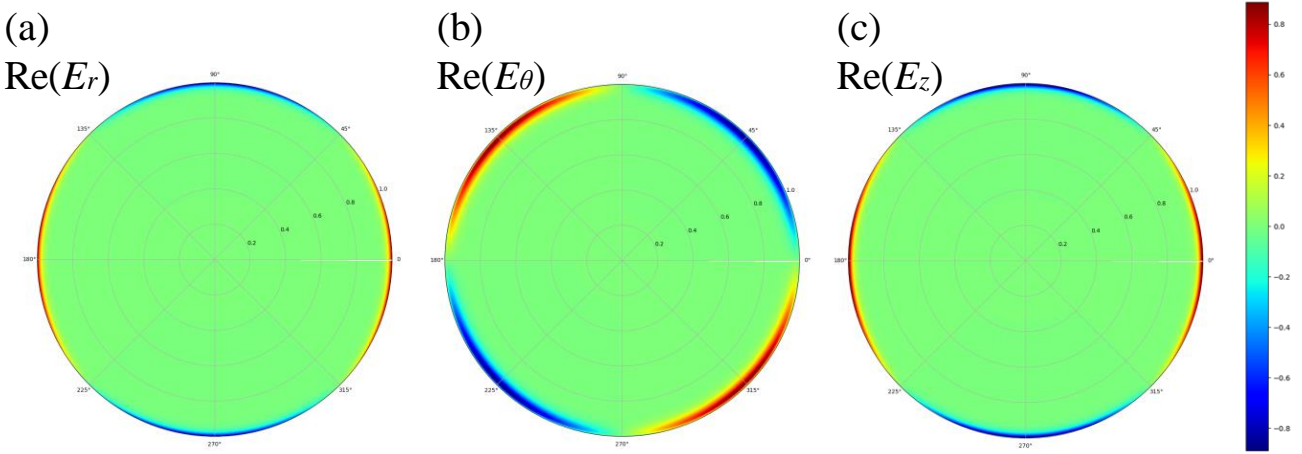

Fig. 2 Field distributions of (a) Re($E_r$), (b) Re($E_\theta$), and (c) Re($E_z$) of plasmonic vortex with $m = 2$ and $R = 10.0$ mm in $r$-$\theta$ plane. All field components of plasmonic vortex are non-zero values. The phase differences among the field components cause the rotations of the polarization, and the rotations produce spin AM in the direction of the rotation axis. The phase difference between $E_r$ and $E_z$ is zero whereas the difference between $E_\theta$ and other components is $\pi/2$.

It is reported that the hybrid polarized surface wave on hyperbolic metasurface carries longitudinal spin AM as well as transverse spin AM [32]. A surface wave has a spin AM being orthogonal to the propagating and decaying directions. In the CCW, the hybrid polarized plasmonic vortex propagates in the axial and azimuthal directions with real axial and azimuthal wavenumbers $k_z$ and $k_\theta = m/R$. The intensity decays in the radial direction with the imaginary radial wavenumber $k_r$. Hence the plasmonic vortex in the CCW would carry axial and azimuthal spin AM which correspond to longitudinal and transverse spin in cylindrical coordinate, respectively.

### III. CALCULATION RESULTS OF SPIN AND ORBITAL ANGULAR MOMENTA

As mentioned above, the total AM density $\boldsymbol{J}$ is expressed as $\boldsymbol{J} = \boldsymbol{r} \times \boldsymbol{p}$. The position vector $\boldsymbol{r}$ is $\boldsymbol{r} = (r, 0, z)$ in cylindrical coordinate. Considering the Floquet's spatial harmonics, the Poynting vector $\boldsymbol{P}$ in the CCW is expressed as the sum of all harmonics as follow.
$$\boldsymbol{P} = \sum_{q=-\infty}^{\infty} Re(\boldsymbol{E_q^*} \times \boldsymbol{B_q})/\mu_0 \qquad (4)$$
The momentum density vector $\boldsymbol{p} = \boldsymbol{P}/c^2$ is decomposed into spin and orbital components as $\boldsymbol{p} = \boldsymbol{p_o} + \boldsymbol{p_s}$ and,
$$\boldsymbol{p_o} = \frac{1}{2\omega}\sum_{q=-\infty}^{\infty} \mathrm{Im}[\epsilon_0(\boldsymbol{E_q^*}\cdot(\nabla)\boldsymbol{E_q}) + \mu_0^{-1}(\boldsymbol{B_q^*}\cdot(\nabla)\boldsymbol{B_q})] \qquad (5)$$
$$\boldsymbol{p_S} = \frac{1}{2}\sum_{q=-\infty}^{\infty} \nabla \times \boldsymbol{S_q} \qquad (6)$$
$$\boldsymbol{S_q} = \frac{1}{2\omega}\mathrm{Im}[\epsilon_0(\boldsymbol{E_q^*} \times \boldsymbol{E_q}) + \mu_0^{-1}(\boldsymbol{B_q^*} \times \boldsymbol{B_q})] \qquad (7)$$
Here, the notation $\cdot(\nabla)$ is Berry's notation [21]. The spin and orbital AM density $s$ and $L$ are expressed as follows.
$$\boldsymbol{J} = \boldsymbol{s} + \boldsymbol{L} \qquad (8)$$
$$\boldsymbol{s} = \boldsymbol{r} \times \boldsymbol{p_s} \qquad (9)$$
$$\boldsymbol{L} = \boldsymbol{r} \times \boldsymbol{p_o} \qquad (10)$$
It should be noted that spin AM density of EM wave are generally expressed as
$$\boldsymbol{S} = \frac{1}{2\omega}\mathrm{Im}[\epsilon_0(\boldsymbol{E^*} \times \boldsymbol{E}) + \mu_0^{-1}(\boldsymbol{B^*} \times \boldsymbol{B})] \qquad (11)$$
because [22, 24]
$$\tfrac{1}{2}\int \boldsymbol{r} \times \boldsymbol{\nabla} \times \boldsymbol{S}\, dV = \int \boldsymbol{S}\, dV. \qquad (12)$$

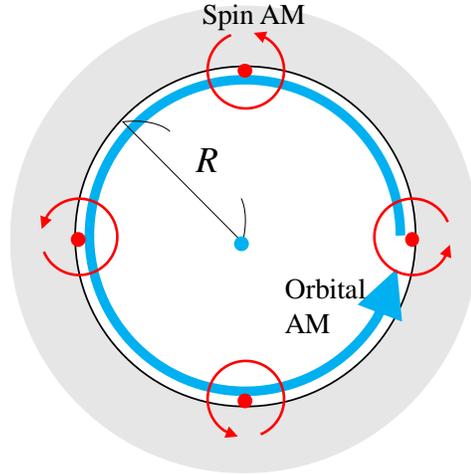

Fig. 3 Blue and red arrows respectively represent rotations of phase and polarization of plasmonic vortex in $r$-$\theta$ plane. The rotations of the phase and the polarization produce the orbital and spin AM, respectively. The rotation axis for the orbital AM is positioned at the center of the CCW (blue dot). On the other hand, the axis for the spin AM is at each point of the vicinity of the wall of CCW (red dots).

Here, $\int dV$ indicates an integration in adequate volume. The spin AM is intrinsic value in this case. In theoretical studies by K. Y. Bliokh et al., a spin AM of a surface wave on a planar surface is analyzed and satisfies equation (12) [18-20]. Equation (12) is valid for any localized fields vanishing at infinity [24]. On the other hand, the plasmonic vortex is confined to the vicinity of the cylindrical wall with the finite radius. The contribution of the position vector cannot be ignored, and then
$$1/2 \int \boldsymbol{r} \times \boldsymbol{\nabla} \times \boldsymbol{S_q}\, dV \neq \int \boldsymbol{S_q}\, dV \qquad (13)$$
with the finite domain of the integration. Hence the spin AM of the plasmonic vortex is expressed by equation (9). In this case, the spin AM is extrinsic value which depends on choice of the coordinate origin.

The orbital AM of the plasmonic vortex is obtained by applying equations (2) and (3) to (10). The axial component of the orbital AM density $L_z$ is
$$L_z = r p_{o\theta}$$
$$= \frac{m}{2\omega}\sum_{q=-\infty}^{\infty}[\epsilon_0\left(|E_{qr}|^2 + |E_{q\theta}|^2 + |E_{qz}|^2\right)$$
$$+ \mu_0^{-1}(|B_{qr}|^2 + |B_{q\theta}|^2 + |B_{qz}|^2)]$$
$$= \frac{m}{\omega}U \qquad (14)$$
Here, $p_{o\theta}$ is azimuthal component of $\boldsymbol{p_o}$, and $U$ is EM energy density of the plasmonic vortex as below.
$$U = \tfrac{1}{2}\sum_{q=-\infty}^{\infty}[\epsilon_0\left(|E_{qr}|^2 + |E_{q\theta}|^2 + |E_{qz}|^2\right) +$$
$$\mu_0^{-1}(|B_{qr}|^2 + |B_{q\theta}|^2 + |B_{qz}|^2)]. \qquad (15)$$
By integrating $L_z$ over the CCW volume, equation (14) is modified as

$$\frac{\int L_z dV}{\int U dV} = \frac{m}{\omega} = \frac{m\hbar}{\omega\hbar} \quad (16)$$

Equation (16) means that the plasmonic vortex in CCW carries orbital AM of $m\hbar$ per photon energy of $\omega\hbar$ which is intrinsic value.

As well as orbital AM, the total AM and the spin AM per photon are calculated by integrating the ratio of AM density to the energy density. The azimuthal components of the AM are negligible because they are much smaller than the EM energy of the plasmonic vortex. Hence, we consider axial spin and total AM in this study. The axial spin AM per photon is obtained by $\int \omega s_z / U dV$, and the axial spin AM density $s_z$ is

$$s_z = -\frac{1}{2} \sum_{q=-\infty}^{\infty} r \frac{\partial S_{qz}}{\partial r}. \quad (17)$$

Here, $S_{qz}$ is axial component of $S_q$. The angular frequency is obtained by the field-matching method, and the frequency is shown in Fig. 4 (a). The calculation results of the spin AM are shown in Fig. 4 (b). The horizontal axis is topological charge $m$. The value of the axial spin AM depends on topological charge, and becomes positive (negative) with the positive (negative) topological charge. The spin AM is caused by the rotation of the polarization in $r$-$\theta$ plane as seen in Fig. 3. The rotation is caused by the phase difference between $E_r$ and $E_\theta$, and the rotation axis is confined at each point of the vicinity of the cylindrical wall with the radius $R$. The axial spin AM is zero with $m = 0$ because the azimuthal phase distribution seen in Fig. 2 vanishes and then the phase difference is always zero. The spin AM of plasmonic vortex coexists with the orbital AM. With sufficiently large orbital AM ($m > 1$), the spin AM is larger than $|\sigma|$ with $m > 1$. The spin AM for $R = 10.0$ mm and $14.85$ mm are almost same from $m = 0$ to $4$, and the spin AM for $R = 14.85$ mm becomes larger than that for $R = 10.0$ mm with $m > 4$. The extrinsic feature causes this $r$-dependence of the spin AM. In Fig. 4 (c) shows axial total AM per photon $\int \omega J_z / U dV$ depending on topological charge $m$. The calculated total AM is equal to the sum of the spin AM in Fig. 4 (b) and the topological charge $m$. The spin and orbital AM are in the same direction, so that the total AM is the sum of the extrinsic spin and the intrinsic orbital AM in agreement with equation (8).

## IV. CONCLUSION

In conclusion, we analyze orbital and spin AM of plasmonic vortex in 0.1-THz CCW. We demonstrate the large extrinsic spin AM of the plasmonic vortex in CCW. The optical extrinsic spin AM is reported in this letter for the first time. The spin AM can be adjusted by the radius of CCW owing to its extrinsic feature. The plasmonic vortex with sufficiently large orbital AM carries the spin AM larger than $|\sigma/\hbar|$ per photon. The large spin AM expands its degree of freedom and may be applied in spin-orbit photonics in terahertz frequency [33]. The plasmonic vortex up to $m = 30$ in the CCW can be excited based on Cherenkov instability, generating intense fields [27, 28]. The spin-orbit coupling in such intense terahertz field would be investigated in our future research.

This work has been partially supported by JSPS KAKENHI Grant Number 20K15191.

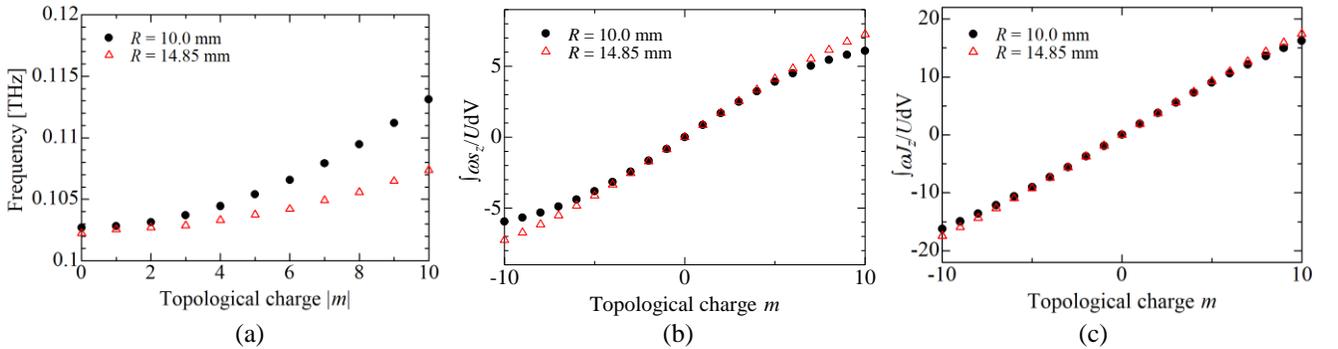

Fig. 4 (a) Frequency of plasmonic vortex at axial wavenumber $k_z = 62.83$ cm$^{-1}$. Calculation results of (b) spin and (c) total AM per photon carried by plasmonic vortex with $k_z = 62.83$ cm$^{-1}$. Total AM per photon is the sum of spin AM of (b) and topological charge $m$ in agreement with equation (8).